\documentclass[%
 reprint,
 amsmath,amssymb,
 aps,
]{revtex4-2}

\usepackage{graphicx}
\usepackage{dcolumn}
\usepackage{bm}

\begin{document}

\preprint{APS/123-QED}

\title{Nucleosome simulations suggest mechanisms of electrostatically-driven mesoscale chromatin evolution}

\author{Siddhartha G. Jena}
 \altaffiliation[]{Harvard Department of Stem Cell and Regenerative Biology, Broad Institute of MIT and Harvard, sidujena@gmail.com}

\date{\today}

\begin{abstract}
Nucleosomes are structures made up of proteins called histones that bind and compact DNA, driving the mesoscale organization of the chromatin polymer. Although histone proteins have diversified over evolutionary time, their contributions to the corresponding diversification of chromatin structure are poorly understood. Here, we mine protein databases for histones and create \emph{in silico} nucleosomes for 3241 organisms spanning $>$1.5B years of evolution. Using a combination of electrostatic calculations and coarse-grained molecular dynamics simulations, we reveal extensive biophysical diversification of the nucleosome unit. Finally, we perform coarse-grained oligonucleosomal simulations on a subset of evolutionarily and biophysically divergent nucleosomes, demonstrating dramatic differences in bulk phase behavior of chromatin. Taken together, our results suggest a paradigm in which histones may have evolved to facilitate particular types of mesoscale chromatin behavior.
\end{abstract}

\maketitle


\section{\label{sec:level1}Introduction}
Chromatin is made up of DNA, RNA, and proteins, the latter of which serve diverse enzymatic, regulatory, and structural functions. While proteins such as transcription factors bind to specific sites in the genome, architectural proteins such as histones are less sequence specific and therefore much more widespread across the genome, their interactions driving the mesoscale spatial organization of the nucleus. While we have gained more appreciation of how cells may have evolved to satisfy developmental, environmental, or physical constraints, we lack an understanding of how diversification in architectural proteins directly maps to this diversification. Recent work has shown that even single amino acid changes or subtle structural alterations in architectural proteins can profoundly reshape the mesoscale organization of chromatin, motivating a more comprehensive map of architectural protein evolution \cite{jena2025engineered}. 
Here, we generate evolutionary-scale databases of histone proteins, and use structural alignment to create a corresponding collection of predicted \emph{in silico} nucleosomes. We show that these nucleosomes have distinct and divergent electrostatic profiles, and that these affect the energetics of internucleosomal interactions. Finally, we use coarse-graining methods to model oligonucleosome chains made up of evolutionarily and biophysically divergent nucleosomes, revealing emergent differences in mesoscale polymer behavior stemming from nucleosome properties and internucleosomal interactions. Together, these results illustrate the ability of histone and nucleosome evolution to shape genome organization across the tree of life, and present \emph{in silico} methods to better understand these trends and potentially inform the design of mesoscale chromatin structures.

\section{\label{sec:level1}Results}
\subsection{\label{sec:level2}An evolutionary-scale database of histone sequences}
To build a structural resource capable of spanning the diversity of
histone proteins across the tree of life, we queried the Pfam database for annotated protein families corresponding to the four canonical core histones (H2A, H2B, H3, H4). We wanted to make sure we were collecting as many histone proteins as possible, including those that may be structurally quite divergent from more well-studied histones. Therefore, in our structural search we additionally included more distantly related histone-fold-containing protein families, including linker histones (H1/H5) and the structurally analogous but evolutionarily distinct histone-fold proteins found in archaea (HMf) and bacteria (HU/IHF), designing our sequence-mining strategy around five Pfam domain families spanning this full range, from the canonical core histone fold to prokaryotic and archaeal
histone-like proteins, as well as the centromere-specific CENP-T
histone fold, providing an additional internal comparison point for a lineage-restricted, functionally specialized histone variant.

For each Pfam family, we used two independent retrieval routes:
direct UniProt cross-reference search, and an InterPro-mediated
accession lookup, rather than relying on either database alone. We
took this approach deliberately after finding that the two databases'
underlying protein classification pipelines do not always agree on
family membership at the margins; querying both and pooling results
was a simple way to reduce the risk of under-sampling a given
Pfam family due to idiosyncrasies of either single retrieval method,
at the cost of producing a combined dataset requiring subsequent
deduplication and quality filtering. 

We retrieved 117,589 sequences corresponding to the core histones, 25,190 sequences corresponding to the linker histones, 333 bacterial histone-like sequences, and 5,056 archaeal histone-like sequences. After removing fragmentary, low-complexity, or premature-stop-codon-containing sequences, we were left with 143,812 total putative histone proteins. This combined sequence database was subsequently curated to the species level and used to select single representative histone sequences (one per core histone) per species, from which to construct template-based homology models of the full histone octamer. The resulting final database contained 3,142 nucleosomes, spanning lineages as far back in evolutionary time as e.g. \emph{Naegleria} and therefore suggesting an approximate evolutionary span of at least 1.5 billion years (\textbf{Figure 1A}).

\begin{figure*}
\includegraphics[scale=1.1]{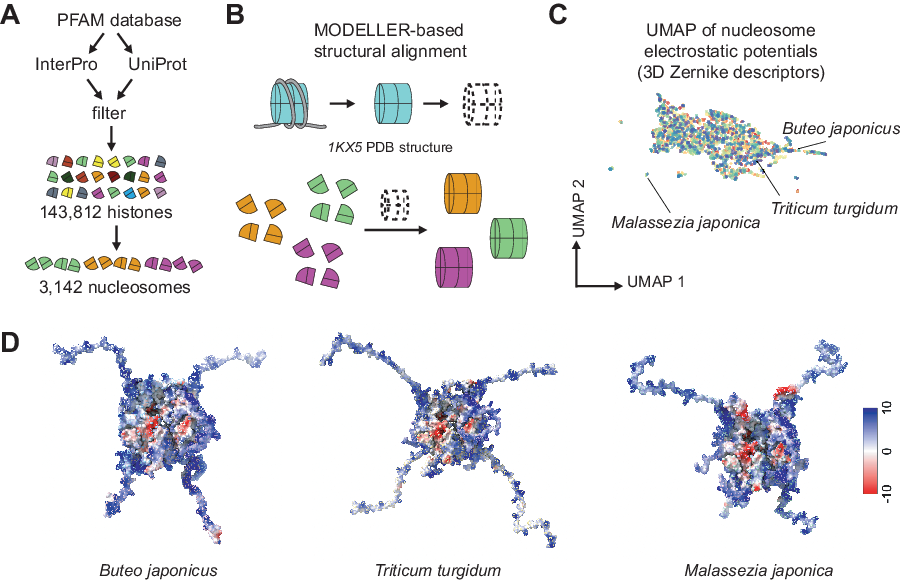}
\caption{\label{fig:wide}A. Schematic of PFAM database mining, filtering, and assignment of 143,812 histones into 3,142 species-specific nucleosomes. B. MODELLER-based structural alignment of organismal nucleosomes onto the 1KX5 structure, allowing for rapid structural assignment and subsequent energy minimization. C. UMAP of nucleosome electrostatic potentials (colored by phyla), calculated using the 3D Zernike descriptors for each nucleosome. Sample outlier nucleosomes are highlighted. D. Structures of nucleosomes highlighted in the UMAP above, surface colored by charge with red corresponding to negative charge and blue to positive charge.}
\end{figure*}

\subsection{\label{sec:level2}\emph{In silico} modeling reveals electrostatic diversity in eukaryotic nucleosomes}
Starting from the ordered histone-fold core of the \emph{Xenopus laevis}
nucleosome core particle (PDB 1KX5) as a single structural template, we
constructed homology models of the full histone octamer for each
species in our sequence database using MODELLER \cite{webb2016comparative}. For each species, a per-species alignment file was constructed, pairing each species' eight histone chains (two copies each of H3, H4, H2A, and H2B) against the corresponding ordered-core template chains from 1KX5, following the standard MODELLER multi-chain alignment format). Since disordered tails can differ significantly between homologs, we used the MODELLER-output B-factors: residues with B-factor $<50$ \AA$^2$ were classified as belonging to the ordered histone-fold core, and residues flanking this ordered region at the N- and C-termini were classified as disordered tails and excluded from the modeling template. N- and C-terminal tails were then added back in after structural alignment. Homology models were built using MODELLER's \texttt{AutoModel} routine
with the trimmed, ordered-core region of 1KX5 as the single known
template structure, generating one model per species (\texttt{starting\_model
= ending\_model = 1}, default MODELLER optimization and refinement
schedule).

Each homology model was energy-minimized in explicit solvent using
GROMACS \cite{van2005gromacs} to relax homology-modeling artifacts
prior to electrostatic calculation. For each species, the model was
placed in a cubic simulation box (minimum 1.0 nm solute-wall clearance),
solvated with explicit water (SPC water model), and neutralized with
Na$^+$/Cl$^-$ counterions to electrical neutrality (\texttt{gmx genion}).
The solvated, neutralized system was energy-minimized using steepest
descent, and the resulting minimized protein coordinates were extracted
(\texttt{gmx trjconv}, with periodic-boundary correction and
recentering) as a solvent-free PDB structure for downstream
electrostatic and structural analysis (\textbf{Figure 1B}). 

Electrostatic potentials were computed with the Adaptive Poisson–Boltzmann Solver (APBS) \cite{holst2000adaptive}, which solves the linearized Poisson–Boltzmann equation on a discretized grid to obtain the solvent-screened electrostatic potential surrounding a macromolecule given its atomic charge distribution. Each structure was first processed with PDB2PQR \cite{dolinsky2007pdb2pqr} to assign partial charges and radii, and per-structure APBS grid parameters (dimensions, center, resolution) were determined automatically for each structure to ensure consistent boundary placement across the full structural database.

Finally, we sought to analyze electrostatic properties of assembled and minimized nucleosomes across thousands of structures. To quantify shape-invariant features of each species' predicted surface electrostatic potential in a form suitable for unsupervised comparison and clustering across the full structural database (independent of each structure's arbitrary orientation in space), we computed rotation-invariant 3D Zernike moment descriptors \cite{novotni20033d} of the volumetric electrostatic potential (EP) field returned by APBS for each homology model. Each APBS grid was first smoothed with a uniform filter and then downsampled, with voxel spacing rescaled accordingly. The downsampled potential field was re-centered on its own charge-weighted centroid. Finally, voxel coordinates were radially normalized to lie within a unit ball, using a length scale set by the 99th percentile of the radial distance among voxels with non-negligible potential magnitude (voxels below 1\% of the field's peak absolute value were excluded from this scale estimate). Voxels falling outside the resulting unit ball were excluded from the moment calculation.

To organize the full structural database of rotation-invariant electrostatic descriptors into an interpretable, low-dimensional representation, we applied a two-stage dimensionality reduction procedure. The Zernike descriptor matrix (one row per species) was first filtered by replacing any non-finite values (arising from structures with degenerate or near-empty potential grids) with bounded finite values, and features with near-zero variance across the database were removed. The remaining features were standardized to zero mean and unit variance, and principal component analysis (PCA) was calculated for the standardized descriptor matrix, retaining a fixed maximum number of components (bounded by the smaller of the configured maximum, the number of retained features, and $N_{\text{species}}-1$). Cumulative explained variance was recorded to assess how much of the descriptor-space variance was captured by the retained components, with 5 components capturing $> 95\%$ of variance (data not shown). The retained principal components were further reduced to a two-dimensional embedding using Uniform Manifold Approximation and Projection (UMAP) \cite{mcinnes2018umap}, with a fixed random seed for reproducibility and neighborhood size, minimum distance, and distance metric set as configuration parameters. This two-stage PCA$\to$UMAP procedure was found to reduce sensitivity to high-dimensional noise in the raw Zernike descriptor space prior to the nonlinear embedding step. The resulting two-dimensional coordinates, together with the retained principal component values were used to generate visualizations of the embedded structural database (\textbf{Figure 1C}).

We next selected, in an unbiased fashion, a tractable subset of species/structures spanning this embedded space for downstream all-atom and coarse-grained characterization. To obtain this subset of structures, we combined three complementary selection criteria:

\emph{1. Cluster representatives.} $k$-means clustering (a configurable number of clusters, $k=10$ by default) was applied to the UMAP embedding, and the medoid structure of each cluster was selected as a representative of that cluster's typical electrostatic phenotype.

\emph{2. Local outliers.} The Local Outlier Factor (LOF) algorithm \cite{breunig2000lof} was applied to the full embedding to rank each structure by the local density of its neighborhood relative to its neighbors' neighborhoods; the highest-ranked structures were selected as the most anomalous points in the embedded space, capturing species whose electrostatic surface is unusual relative to the rest of the database.

\emph{3. Embedding-axis extrema.} The structures with the minimum and maximum coordinate along each UMAP axis.

The union of these three criteria (with duplicate selections resolved by retaining the first-assigned reason) produced the final set of species selected for detailed biophysical characterization (3 representative species annotated in \textbf{Figure 1C}). We selected 12 species total, spanning mammals (\emph{Urocitellus parryii, Callorhinus ursinus, Sus scrofa}), a reptile (\emph{Platysternon megacephalum}), birds (\emph{Oriolus oriolus, Buteo japonicus, Rhipidura dahli, Cyanoderma ruficeps}), plants (\emph{Triticum turgidum}), a nematode (\emph{Globodera pallida}), and fungi (\emph{Lentinus tigrinus, Malassezia japonica}). Importantly, this diversity stems organically from the unbiased methods used to select these structures. Examining electrostatic profiles of all-atom structures for these nucleosomes revealed clear differences in charge distribution across both the globular domain and disordered tails (examples shown in \textbf{Figure 1D}), further highlighting the utility of this approach.

\subsection{\label{sec:level2}Energetic calculations reveal divergence in internucleosomal interactions}

For each species, an \emph{in silico} nucleosome structural model and
associated surface electrostatic potential were generated using the
DISCO pipeline \cite{beard2001modeling}. Each model consists of (i) a set of
$M=1000$ discretized point-charge sites $\{\mathbf{r}_i, q_i\}$
distributed over the nucleosome surface, derived from the previously calculated Poisson-Boltzmann
electrostatic calculations (APBS) on the corresponding histone-octamer
structure, and (ii) an associated geometry file encoding the nucleosome
disk-normal axis and physical dimensions. Point-charge coordinates were
re-centered on their own center of mass prior to all downstream
calculations.

Across the twelve species examined in detail, the net charge of the
DISCO-derived nucleosome charge distribution was positive in every
case, ranging from $27.7\,e$ (\emph{Globodera pallida}) to $141.6\,e$
(\emph{Buteo japonicus}): a five-fold range in magnitude (data not shown). This is consistent with these charge
distributions representing the histone protein surface, which is
strongly enriched in basic (lysine/arginine) residues, rather than the
combined histone-DNA assembly; DNA's phosphate backbone,
which was ignored in this coarse-grained representation,
is the dominant source of net negative charge in the biological
nucleosome. Because every species carries the same charge sign, simple monopole electrostatics predicts uniform repulsion between any two nucleosomes of the same species at long range, regardless of identity.

To test whether net-positive nucleosomes could nonetheless interact
attractively at short range, we computed the full pairwise
Debye--H\"uckel and Weeks-Chandler-Andersen (WCA) interaction energy between two identical copies
of each species' nucleosome, across a grid of center-to-center
separations (60--300~\AA) and relative clocking angles in a co-axial,
face-to-face stacking geometry, a measurement we called a dimer scan (\textbf{Figure 2A}). Despite uniform
net-positive charge, five of twelve species (\emph{Buteo japonicus},
\emph{Cyanoderma ruficeps}, \emph{Platysternon megacephalum}, \emph{Sus
scrofa}, and \emph{Urocitellus parryii}) showed a net-attractive
global energy minimum in this geometry, with well depths ranging across several orders of magnitude (kcal/mol) with \emph{Platysternon megacephalum} displaying the largest attractive energetic contribution. Bulk charge was insufficient to predict dimer interactions; notably, \emph{Buteo japonicus} carried the largest
net charge of any species examined ($141.6\,e$) yet showed only modest attraction, while \emph{Platysternon megacephalum}
with a comparatively unremarkable net charge ($49.0\,e$), showed the
deepest well of all twelve species (representative trends in \textbf{Figure 2B}). 

To resolve the mechanism underlying this attraction and to accurately capture the ideal geometry for each dimer, we decomposed the total interaction energy at each species' best co-axial pose into per-site contributions, summing each charge site's interaction with all 1000 charge sites on the partner nucleosome. This approach was quite enlightening; for instance, in the case of \emph{Platysternon
megacephalum}, a single charge site contributed
$-0.80$~kcal/mol of the total $-1.06$~kcal/mol interaction energy: more than 75\% of the net attractive energy arising from one of 1000
sites, with the standard deviation of all other per-site contributions
more than an order of magnitude smaller ($\sigma = 0.028$~kcal/mol). By
contrast, the same decomposition for \emph{Rhipidura dahli}, evaluated at the identical 107.19~\AA\ separation
for direct comparison, showed a maximum single-site contribution of
only $-2.4\times10^{-4}$~kcal/mol, roughly 3,000-fold weaker than
\emph{Platysternon megacephalum}'s dominant site. The magnitude of the single most attractive site-pair interaction spanned
more than four orders of magnitude across species (from
$3.7\times10^{-8}$ to $0.80$~kcal/mol) (\textbf{Figure 2B}). We interpret this as
evidence that nucleosome-nucleosome attraction in this system is
governed by a small number of spatially localized, high-magnitude
charge features, or "hotspots," rather than by the charge distribution as a whole.

Despite the near-universal presence of an attractive
configuration, the fraction of solid angles corresponding to net
attraction was small for every species: $0.65\%$
(\emph{Platysternon megacephalum}) and $0.025\%$ (\emph{Buteo japonicus}) at their respective optimal separations, with most other
species falling between $0.01\%$ and $3.2\%$ (\textbf{Figure 2C}). This suggests that even where an attractive hotspot exists, it occupies only a narrow window of the full space of relative orientations available to two unconstrained rigid bodies, rather than a broadly interactive surface, consistent with previous treatments of this phenomenon \cite{korolev2018systematic, kepper2008nucleosome}. Finally, comparing energetic contributions to the fraction of time spent in a net-attractive dimer state suggested a weak trend towards energetic biasing of nucleosome-nucleosome interactions in dimers (\textbf{Figure 2D}).

\begin{figure*}
\includegraphics{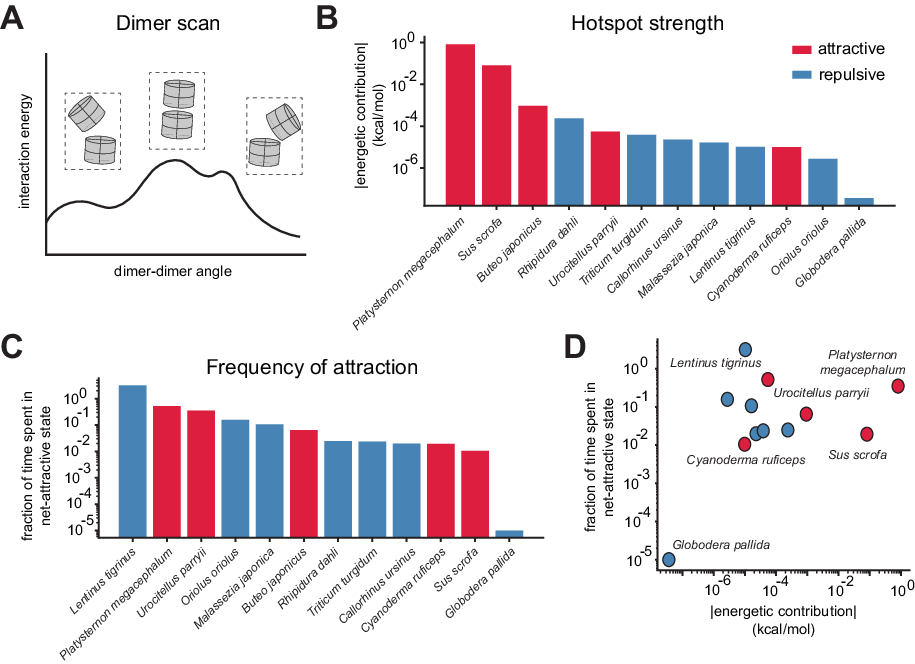}
\caption{\label{fig:wide}A. Schematic of nucleosome-nucleosome dimer scan: the relative orientation of two nucleosome "monomers" is changed and energy calculated at each orientation. B. Absolute value of energetic change at hotspot angles for representative nucleosomes, with attractive interactions in red and repulsive interactions in blue. C. Frequency of attractive interactions (out of all sampled angles) for representative nucleosomes. D. Relationship between energetic contribution of hotspot angles and the fraction of time spent in an attractive state.}
\end{figure*}

\subsection{\label{sec:level2}A patchy-particle model captures hotspot geometry and enables tractable many-body simulation}
We sought to translate the geometrically sensitive energetics captured in our dimer scan to a model that could be extended to large collections of particles in order to tractably simulate the chromatin polymer. We decided to use a patchy-particle system, which allows for such geometrically restricted interactions to be encoded in an otherwise coarse-grained system \cite{ancona2022simulating, arabzadeh2025chromrec, zhang2004self} (\textbf{Figure 3A}).

We extracted an effective single-patch description for each species from the dense full-orientation grid, evaluated at a common separation (107.19~\AA) for cross-species comparability: a patch well depth (the grid's global minimum energy), an angular half-width (back-calculated from the measured attractive solid-angle fraction under a two-independent-patch alignment approximation), and a patch direction (the unit vector in each nucleosome's own body frame, pointing toward its partner at the optimal configuration). In one example, \emph{Platysternon megacephalum}, this
procedure yielded a well depth of $1.26$~kcal/mol ($2.1\,k_BT$) and a
patch half-angle of $28.2^\circ$.

Using these species-specific patch parameters in a Kern-Frenkel-style
patchy hard-sphere Monte Carlo model \cite{kern2003fluid}, we simulated $N=100$ particles along a chain under canonical (NVT) Metropolis sampling (\textbf{Figure 3B}). At a
representative dilute density ($\rho = 2.33\times10^{-6}$~\AA$^{-3}$,
corresponding to a mean interparticle spacing of $\sim$75~\AA), \emph{Lentinus tigrinus} particles displayed increased clustering, aligning within the simulation box (particles with vectors designating patch alignment shown in \textbf{Figure 3C}). However, \emph{Platysternon megacephalum} particles showed only transient pairwise
association, but no higher-order
clustering. This suggests that although \emph{Lentinus tigrinus} interactions were overall repulsive (\textbf{Figure 2B}), the frequency of attractive interactions was sufficient to drive mesoscale clustering behavior (\textbf{Figure 2C}). Finally, comparing the geometry of patch interactions to clustering fraction revealed a trend between patch half-angle and clustering fraction, suggestive of both geometry and electrostatics driving mesoscale organization (\textbf{Figure 2D}).

\begin{figure*}
\includegraphics[scale=0.6]{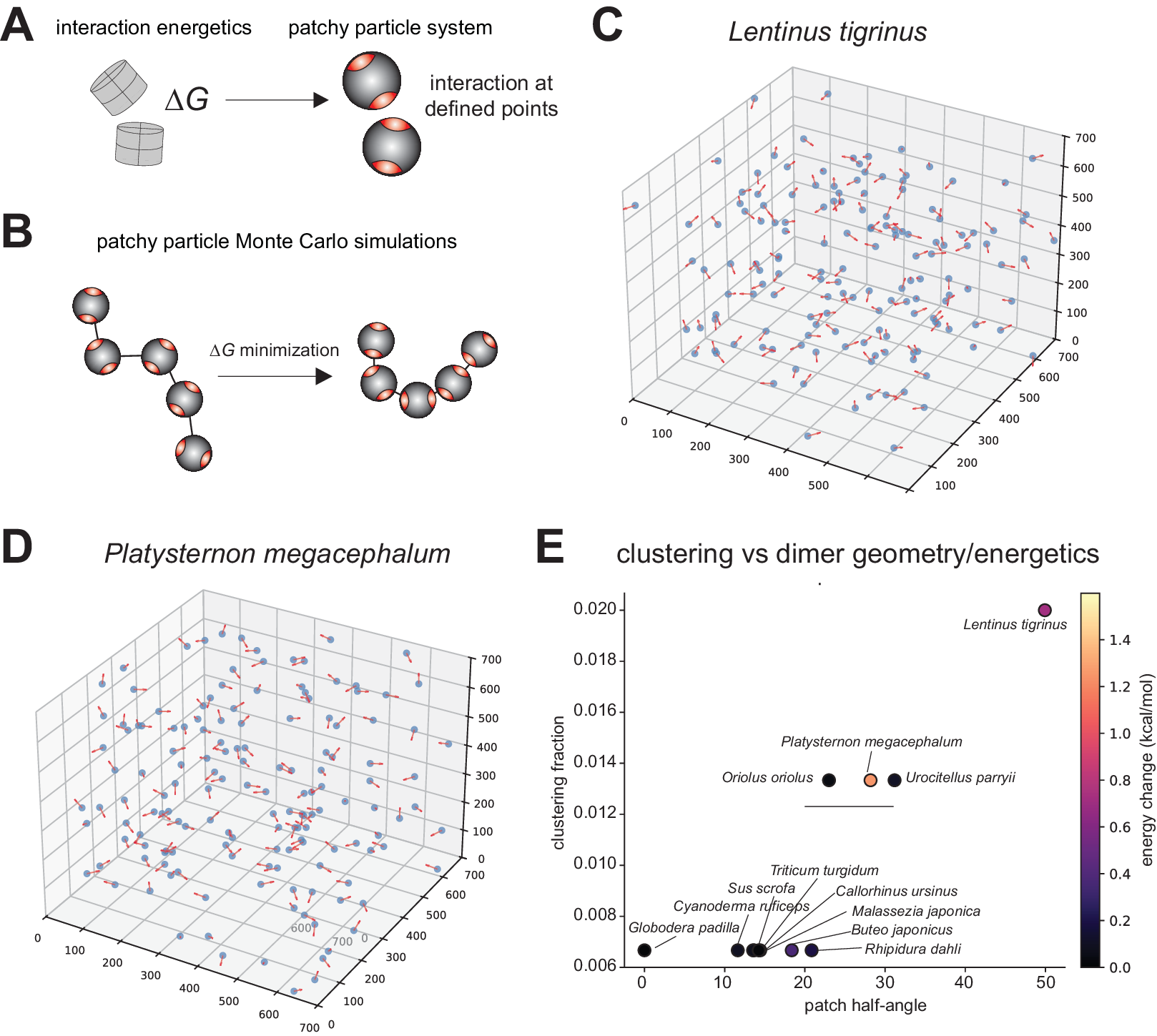}
\caption{\label{fig:wide}A. Schematic of nucleosome-nucleosome dimer scan: the relative orientation of two nucleosome "monomers" is changed and energy calculated at each orientation. B. Absolute value of energetic change at hotspot angles for representative nucleosomes, with attractive interactions in red and repulsive interactions in blue. C. Frequency of attractive interactions (out of all sampled angles) for representative nucleosomes. D. Relationship between energetic contribution of hotspot angles and the fraction of time spent in an attractive state.}
\end{figure*}

\section{\label{sec:level1}Discussion}
We have shown that the electrostatic surface of the nucleosome varies substantially in a manner that is not captured by simple aggregate descriptors such as net charge or dipole moment, but
is instead dominated by the strength and placement
of localized charge hotspots. These hotspots occupy only a small
fraction of the full relative-orientation space available to two
freely diffusing nucleosomes, rendering them thermodynamically
negligible for an unconstrained particle population, but potentially decisive for a polymer-constrained
one, where linker DNA geometry biases nucleosome pairs toward the
narrow orientational window in which the hotspot can engage. By turning this interaction into
an effective anisotropic pair potential, we obtained a coarse-grained
representation capable of directly testing many-body clustering
behavior at scale.

These results support a model in which histone evolution
could tune the mesoscale phase behavior of chromatin not by altering
the nucleosome's bulk charge budget, but by sculpting the presence and
geometry of specific, localized surface features. This reframes the question of how histones encode chromatin organization in a way that is experimentally
and computationally tractable: rather than searching for global
sequence or charge signatures that correlate with compaction phenotype
across species, our results motivate a search for the specific,
localized surface residues responsible for individual hotspots, and a
test of whether engineering or removing them is sufficient to shift a
given nucleosome's mesoscale behavior in a predictable direction. More
broadly, the analysis pipeline developed here, linking structural
databases, Poisson-Boltzmann electrostatics, dimer-level energy
landscapes, and coarse-grained
many-body simulations, provides a general framework for
connecting evolutionary sequence variation in histones to
the emergent physical organization of the genome, and offers a
concrete starting point for the rational design of synthetic chromatin
with tunable mesoscale properties \cite{jena2025engineered}.

Several limitations of the current approach are worth noting. First,
our coarse-grained charge representation collapses the full
conformational and chemical complexity of the histone octamer surface
(including post-translational modifications, which are entirely absent
from this model) into a static point-charge distribution computed from
a single structural model per species; real chromatin electrostatics
are almost certainly modulated dynamically by tail modification state,
which we have not addressed here. Second, our patchy-particle model
uses a single dominant patch per nucleosome, whereas the underlying
atomistic energy landscape may in principle support multiple,
weaker, or context-dependent attractive features that a single-patch
Kern-Frenkel approximation cannot capture.
Finally, we modeled linker DNA and screening electrolyte implicitly
(via a fixed Debye length and dielectric constant); explicit treatment
of linker histones, multivalent cations, and RNA/protein binding
partners known to modulate chromatin phase behavior \emph{in vivo}
represent natural extensions of this framework.

\section{\label{sec:level1}Methods}

\subsection{Dimer scans}

Nucleosome A was fixed at the origin with its disk-normal vector $R\hat{n}_A$ aligned
to the $z$-axis (via the rotation minimizing $\lVert R\hat{n}_A -
\hat{z}\rVert$). An identical copy, nucleosome B, was placed along $+z$
at center-to-center separation $d$ and rotated about the shared $z$-axis
by a ``clocking'' angle $\theta_c$. Total interaction energy was calculated as the sum of Debye--H\"uckel and Weeks-Chandler-Andersen (WCA) energies:
$U_{\mathrm{total}}(d,\theta_c) = U_{\mathrm{DH}} + U_{\mathrm{WCA}}$ and was
evaluated on a grid of separations ($d = 62$--$300$ \AA) and clocking
angles ($\theta_c \in [0^\circ, 360^\circ)$, 24--36 values), for each of
the twelve species. This restricted geometry probes face-to-face
nucleosome stacking, the configuration most consistent with the
short-linker topological constraints of an oligonucleosome chain.

To assess the accessibility of attractive electrostatic configurations
across the complete relative-orientation space of two rigid bodies
(rather than the restricted co-axial family above), a dense grid was
constructed over the polar and azimuthal angle of nucleosome B's
approach direction ($\theta \in [0,\pi]$, $\phi \in [0, 2\pi)$) and its
clocking rotation about that approach axis ($\theta_c \in [0, 2\pi)$),
each discretized into 24 values ($24^3 = 13{,}824$ configurations per
species per separation). For each species, this grid was evaluated at
that species' own global-minimum-energy separation identified from the
co-axial stacking scan, ensuring the orientation-space
survey was centered on each species' most favorable approach distance
rather than an arbitrary shared value. The fraction of solid angle
corresponding to net-attractive configurations ($U_{\mathrm{total}} <
0$) was computed using proper $\sin\theta$ solid-angle weighting,

\begin{equation}
f_{\mathrm{attractive}} = \frac{\sum_{k:\,U_k<0} \sin\theta_k}
{\sum_k \sin\theta_k}.
\end{equation}

Patch parameters for the coarse-grained many-body model
were subsequently extracted from an analogous dense grid
evaluated at a single separation ($d = 107.19$ \AA, the deepest global
minimum observed across all species in the co-axial scan) to enable
direct cross-species comparability of the extracted patch geometry.

\subsection{Hotspot decomposition and patch parameter extraction}
For each species, the per-site contribution to the total interaction
energy was computed at a fixed, cross-species-comparable separation
($d=107.19$ \AA, corresponding to the deepest global minimum observed
across the co-axial scan) as
$c_i = \sum_j U_{\mathrm{DH}}(r_{ij})$, summed over all partner sites $j$
on the opposing nucleosome. Site-pair contributions were ranked to
identify dominant attractive contacts, and the spatial distribution of
$c_i$ was visualized by projecting charge-site coordinates onto the
plane perpendicular to the disk-normal vector.

To parametrize an effective anisotropic (``patchy-particle'') potential
for each species, the following quantities were extracted from the
dense full-orientation grid:
\begin{itemize}
\item \textbf{Well depth} $\epsilon_p$: the magnitude of the global
minimum energy found across the grid.
\item \textbf{Patch half-angle} $\theta_p$: back-calculated from the
measured attractive solid-angle fraction $f_{\mathrm{attractive}}$ under
the approximation that two independent single-patch particles must both
achieve alignment, $f_{\mathrm{attractive}} \approx f_{\mathrm{particle}}^2$,
giving
$f_{\mathrm{particle}} = \sqrt{f_{\mathrm{attractive}}}$ and
$\theta_p = \arccos(1 - 2f_{\mathrm{particle}})$
(standard spherical-cap solid-angle relation).
\item \textbf{Patch direction} $\hat{p}$: the unit vector, in each
nucleosome's own body frame, pointing toward the energy-minimizing
configuration of its partner.
\end{itemize}
Self-consistency of the extracted patch direction was assessed via the
cosine similarity between the patch directions independently recovered
from nucleosome A's and nucleosome B's frames at the same optimal
configuration (expected $\approx 1$ for a genuine single, localized,
mutually-interacting feature).

\subsection{Patchy-particle Monte Carlo simulations}
Many-body clustering behavior was assessed using a Kern--Frenkel-style
patchy hard-sphere model \cite{kern2003fluid}, parametrized
per-species by the well depth, patch half-angle, and (nominal) patch
direction extracted above. Each particle possesses an isotropic WCA
repulsive core (species-specific $\sigma_c$) and a single attractive
patch. Two particles $i,j$ interact attractively, in addition to the
always-active WCA repulsion, only when (i) the inter-particle unit
vector lies within patch $i$'s half-angle cone, (ii) the reciprocal unit
vector lies within patch $j$'s half-angle cone, and (iii) the
center-to-center distance lies within the attractive well range
$[\sigma_c, \sigma_c + 30\text{ \AA}]$; under these conditions the pair
potential is offset by $-\epsilon_p$ (square-well approximation).

Simulations of $N=100$ particles in a periodic cubic box were performed
using canonical (NVT) Metropolis Monte Carlo with combined
translational (max.\ displacement 15 \AA) and rotational (max.\ angle
$25^\circ$ about a random axis) trial moves, accepted with probability
$\min(1, e^{-\Delta U/k_BT})$. Number density was varied by scanning box
length $L \in \{250, 350, 450, 550, 700\}$ \AA\ at fixed $N$. Systems
were evolved for $3000$ MC sweeps (one sweep = $N$ attempted moves per
particle); clustering order parameters were evaluated from the final
configuration using a bonding-criterion adjacency graph (connected
components under the same alignment-and-distance bonding rule used to
define attractive interactions, distinct from a naive distance-only
cutoff, which was found to conflate geometric crowding with genuine
attractive bonding at higher densities). Reported observables include
the largest-cluster fraction, fraction of particles with at least one
active bond, cluster-size distribution, and radial distribution
function $g(r)$.

\subsection{Trajectory and configurational analysis}
For oligonucleosome chain trajectories, the following observables were
computed from nucleosome core-bead coordinates: radius of gyration
$R_g$ and end-to-end distance $R_{ee}$; the gyration tensor and derived
shape anisotropy $\kappa^2 = \tfrac{3}{2}\sum\lambda_i^2/(\sum\lambda_i)^2
- \tfrac{1}{2}$; mean pairwise-distance and contact-probability maps
(contact threshold 120 \AA, excluding nearest-neighbor pairs); bond,
angle, and dihedral distributions; tangent--tangent correlations fit to
a worm-like-chain decay to estimate an apparent persistence length;
center-of-mass and internal mean-squared displacements; autocorrelation
functions and integrated autocorrelation times for $R_g$, $R_{ee}$, and
contact number; Rouse-mode decomposition; and PCA/$k$-means
conformational clustering (cluster number selected by silhouette score).

\subsection{Software and statistical analysis}
Molecular dynamics simulations were performed in LAMMPS (stable release). All downstream analyses were
implemented in Python 3 using NumPy, SciPy, pandas, scikit-learn, and
Matplotlib.


\bibliography{apssamp}

\end{document}